\documentclass[%
reprint,
amsmath,amssymb,
aip,
jcp
]{revtex4-2}

\usepackage{graphicx}
\usepackage{dcolumn}
\usepackage{bm}

\usepackage[version=3, arrows=pgf-filled]{mhchem} 

\begin{document}
	\title{Near-Ambient Pressure Velocity Map Imaging}
	
	\author{Tzu-En Chien}
	\affiliation{Dept. of Chemical Engineering, KTH Royal Institute of Technology, Stockholm 100 44, Sweden} 

	\author{Lea Hohmann} 
	\affiliation{Dept. of Chemical Engineering, KTH Royal Institute of Technology, Stockholm 100 44, Sweden} 

	\author{Dan J. Harding}
	\email{djha@kth.se}
	\affiliation{Dept. of Chemical Engineering, KTH Royal Institute of Technology, Stockholm 100 44, Sweden} 
	
	\date{} 

\begin{abstract}

We present a new velocity map imaging instrument for studying molecular beam surface scattering in a near-ambient pressure (NAP-VMI) environment. The instrument offers the possibility to study chemical reaction dynamics and kinetics  where higher pressures are either desired or unavoidable.
NAP-VMI conditions are created by two sets of ion optics that guide ions through an aperture and map their velocities. The aperture separates the high pressure ionization region and maintains the necessary vacuum in the detector region. The performance of the NAP-VMI is demonstrated with results from \ce{N$_2$O} photodissociation and \ce{N$_2$} scattering from Pd(110) surface, which are compared under vacuum and at near-ambient pressure ($ 1\times10^{-3} $ mbar). NAP-VMI has the potential to be a applied to, and useful for,  a broader range of experiments including photoelectron spectroscopy and scattering with liquid microjets. 
\end{abstract}

\maketitle

\section*{Introduction}

Particle imaging techniques  based on Ion Imaging\cite{chandler1987two} and Velocity Map Imaging (VMI)\cite{eppink1997velocity} have revolutionized the field of chemical dynamics, allowing very efficient collection of the velocity distributions of particles coming from a variety of processes. (photodissociation,\cite{chandler1987two, eppink1997velocity, Suits2018Photofragment}  photoelectron spectroscopy,\cite{eppink1997velocity, Chatterley14time, Weichman2018SEVI} inelastic\cite{Vogels2015Imaging, Heid2019Side, deJongh2020Imaging} and reactive scattering\cite{Pan2020Rainbow}) Charged particle-neutral collisions can also be probed\cite{Carrascosa2017Imaging, Meyer2021Atomistic, Bull2013Quantification} The velocity distributions provide information about the physics underlying and controlling the process of interest. The high collection efficiency is particularly valuable for time-resolved experiments, where many measurements need to be made, and for low count rate experiments.

Imaging techniques are still in very active development. Hardware is being pushed in different directions to give ion optics with higher energy resolution\cite{eppink1997velocity,Wrede2001, plomp2021velocity},  larger energy acceptance,\cite{Horke2012VMI,skruszewicz2014new, kling2014thick, schomas2017compact}  or smaller sizes.\cite{sakkoula2021compact} 
Efforts are also ongoing to develop faster, more flexible, and more accurate image reconstruction methods. \cite{manzhos2004photofragment, POP, FINA, PyAbel, AIR22}



Many of these developments are related to the fact that during the imaging detection process a 3D-distribution of particles is projected onto a 2D-detector. There are different approaches to deal with this, depending largely on the nature of the experiment. The most common way, if the experimental set--up has suitable symmetry, is to apply an inverse-Abel transform to the 2D image to recover the original 3D distribution. For systems without this symmetry there are several options including: 
3D imaging where the position and arrival time of each particle is measured,\cite{chichinin2009imaging} recently extended to electrons by the use of the TPX3CAM fast camera.\cite{cheng20223d}
Time slicing approaches, which use more complex electric fields to spread the ions' arrival times at the detector allowing only a chosen section of the ion distribution to be imaged.\cite{gebhardt2001slice, townsend2003direct, lin2003application, papadakis2006slice,ryazanov2013improved, allum2021post}
There are also options to spatially select the particles that are detected by laser slicing, where only a subset of the scattering products are ionized,\cite{TONOKURA1994Slicing} or space slice imaging, where the ions are selected with physical slit before being accelerated perpendicularly on to a detector.\cite{mizuse2019space} 

Additional optical elements have also been used to zoom in or out on images. In particular, Offerhaus et al., used an einzel lens as a magnifier for PEI,\cite{offerhaus2001magnifying} which had a hard focus, or crossover, similar to that reported here. Due to the different application, it was much further from the extraction region.

These strengths of imaging detection have recently been applied to molecular beam surface scattering experiments, where they can be used to probe the dynamics of the molecule-surface interactions\cite{harding2015using, Greenwood2021Nitric} Using pulsed molecular beams and lasers even allows pump-probe measurements of the kinetics of surface processes and reactions by directly probing the time-dependent flux of molecules leaving the surface.\cite{harding2017ion, Neugebohren2018velocity, DimaNH321}  Wodtke and co-workers are developing high repetition rate imaging experiments that can probe the time dependence within a single MB pulse, and will allow changes in the surface during reaction to be investigated.\cite{DimaTR20}

Typically, the pressure in the extraction region of an imaging system must be kept in the high vacuum region (below $1\times10^{-6}$ mbar) to avoid collisions of the particles with background gas and to prevent electrical arcing on the ion optics, or worse, the microchannel plate(s) and phosphor screen of the detector. 
For experiments on gas phase molecules these pressure requirements do not typically pose a problem and the normal solution of well defined molecular beams also improve other aspects of the experiment (e.g. energy resolution in scattering experiments, well defined interaction region). 
When studying gas-solid or gas-liquid interfaces there are many cases where it is not possible or desirable to work in such high vacuum, either because the condensed phase has too high a vapor pressure (e.g. volatile liquids) or because the surface is changed by the presence of reactive gases. 
This has lead to the development of instrumentation to allow `high-vacuum' techniques to be applied at higher pressures. The names given to these different adaptations depend upon the field, but in general use some combination of confining the high pressure region while allowing the particles of interest to reach a detector without losing the useful information they carry.
The surface science community has long recognized the importance of the reactant gas pressure in changing the structure and properties of surfaces, known as the `pressure gap' between ultrahigh vacuum (UHV) and applied catalysis.\cite{lundgren2017novel, van2017surface, newton2017time}  In order to help close the gap it is also important to be able to switch rapidly between UHV and high(er) pressure.

We will briefly review some areas where surface methods detecting charged particles have been extended to higher pressure and which are closely related from their technology and/or applications to the NAP-VMI we describe here.

One of the most widely used techniques for fundamental catalysis research is photoelectron spectroscopy (PES), and there have been significant efforts starting in the 1970s and taking off in the early 2000s to allow PES to be performed at higher `near-ambient'  and ambient pressures.\cite{starr2013investigation, schnadt2020present} In the early instruments small apertures and differential pumping were used to separate the interaction region from the analyzer and maintain suitable vacuum. The use of electron lenses in the differential pumping region allowed higher transmission and pressures.\cite{ogletree2002differentially} The addition of reaction cells\cite{knudsen2016versatile}
and, most recently, developments using virtual cells have pushed the achievable surface pressure above 1 bar.\cite{amann2019high}
Time resolved APXPS experiments have also been developed \cite{papp2013situ, redekop2021synchronizing} and experiments with time resolution of 50 ms have very recently been reported.\cite{knudsen2021stroboscopic}
AP-XPS instruments also provide means to measure the gas composition in front of the surface, with XPS, and in the bulk gas mixture, with mass spectrometer(s), which allows the surface composition to be linked to the reactivity. Care must be taken in using mass spectrometry (MS) as comparison of NAP-XPS, Planar Laser Induced Floresence (PLIF) and MS shows that the composition directly in front of the surface is different to the average composition in the reactor, which is what is sampled by MS in most cases.\cite{blomberg2016comparison,  schnadt2020present}  
These features have allowed AP-XPS to make huge advances in determining the chemical species that are present during reactions at elevated pressure and identifying which are the most reactive. The time resolved measurements are moving towards direct measurement of the surface reaction kinetics but, to our knowledge, these have not been reported so far.

Liquid-gas interfaces are another important area for chemistry where high pressures of gas are often unavoidable, due to evaporation of the liquid. 
Experiments to probe the scattering and evaporation of molecules from liquid surfaces, i.e. measurements of the residence times of atoms and molecules on the surface and/or velocity distributions with which they leave, can provide information about the processes occurring at the interface and the liquid surface composition and structure.\cite{tesa2016atomic, smoll2019scattering} 
For low vapor pressure liquids, a partially submerged wheel rotating through the liquid can be used to produce a continuously clean surface. Techniques similar to those used for gas-solid scattering can then be used.\cite{nathanson2004molecular,zutz2017angle,smoll2018probing, bianchini2019real} Recently, VMI has been also applied to MB scattering from SAMs on metal surfaces.\cite{roscioli2011quantum, hoffman2016quantum}
Liquid metals can also have sufficiently low vapor pressure to be studied with MB methods similar to those for the solids, providing information about energy transfer processes.\cite{zutz2020nonadiabatic}
Liquid microjets\cite{faubel1988molecular} have made it possible to study evaporation from volatile liquids\cite{faubel1988molecular, ryazanov2019quantum} and molecular beam scattering from salty water.\cite{faust2016gas}

The Environmental Molecular Beam (EMB) technique, which has been developed in Gothenburg to allow scattering experiments from volatile liquids, uses an aperture or a grating to separate the high pressure surface from the low pressure detector region.\cite{kong2012ice,johansson2017novel} This allows high pressure and very short flight distances of the molecules through the high pressure region. The use of electron ionization provides universal detection but prevents quantum state specific measurements. Compared to imaging detection, this design can measure only one angle at a time and the data analysis is more complicated, particularly to separate surface residence time distributions from the molecular speed distribution.\cite{Johansson19Understanding} 

As for solid surfaces, photoelectron spectroscopy can provide information about the liquid surface.
The gas load in the system can be reduced by the use of liquid microjets\cite{faubel1988molecular}\cite{faubel1997photoelectron} which, when combined with AP-XPS analyzers, allow photoelectron spectroscopy of liquid interfaces.\cite{seidel2016valence, dupuy2021core}
Developments in the area are being driven by the desire to gain more information from the emitted photelectrons. Interest in the angular distribution of the slow potoelectrons, particularly for measurements of circular dichroism, has led to the development of an instrument combining a liquid jet with movable analyzer\cite{malerz2022setup} and a very recent report of a VMI set up to probe the angular distribution.\cite{long2021liquid}  
Tesa-Serate et al.,\cite{tesa2016atomic} suggested that more `more sophisticated imaging methods for detecting the scattered products' will be useful for molecular beam liquid surface experiments.



Here, we present a new instrument developed to allow velocity map imaging of molecular beam (MB) surface scattering and reactions occurring at near-ambient pressure (NAP-VMI). 
Very high resolution velocity is not a requirement for imaging detection to be useful for surface scattering experiments as, even starting with very well defined beams e.g. ref \cite{Buenermann2015Electron}, most surface scattering processes from metals produce rather broad distributions. Despite this, dynamic fingerprints can still be observed in scattering distributions which can help to separate different processes or reaction channels.\cite{Neugebohren2018velocity, Jiang2019Imaging} We expect this to be the case for the reactive scattering we intend to study.


We will describe the NAP-VMI ion optics 
and show tests of the set up using the well characterized photodissociation of gas-phase  N$_2$O, and MB scattering of N$_2$ from Pd(110) at pressures up to $1\times10^{-3}$ mbar. 
The NAP-VMI can be run in two different modes of operation, one as a `classical' three electrode VMI\cite{eppink1997velocity} and a second mode, using two sets of electrodes with a crossover of the ions at the aperture which provides DC slicing\cite{townsend2003direct,lin2003application}. This mode has the potential to allow even higher pressures by allowing smaller apertures, lower extraction voltages in the high pressure region, and additional stages of differential pumping by refocusing the particles through more apertures, similar to many current AP-XPS analyzers. 
	

Our NAP-VMI instrument enables unique state-resolved gas-solid MB scattering experiments and we suggest that the method has the potential to be interesting and useful to a much broader range of the chemical dynamics community. 
The acceptable pressure in interaction region should allow MB scattering from volatile liquid surfaces  
or liquid jets  
and, as Long et al.\ very recently demonstrated, angular information from liquid jet PES. \cite{long2021liquid}  
Some of the features of our implementation may be particularly useful for LJ-PES and LJ-PECD, in particular, the small aperture size would allow the use of flatjet, providing a better defined system for angular measurements\cite{Malerz2022EASI} and allow all photoelectrons to be probed simultaneously.  
In the crossing mode, the low voltage on the first electrode stack might also reduce the problems with electric breakdown that Long et al. describe for their implementation.\cite{long2021liquid}




\section*{Methods}

\subsection*{A. Overall design} 
\begin{figure}[h]
    \centering
    \includegraphics[scale=0.125]{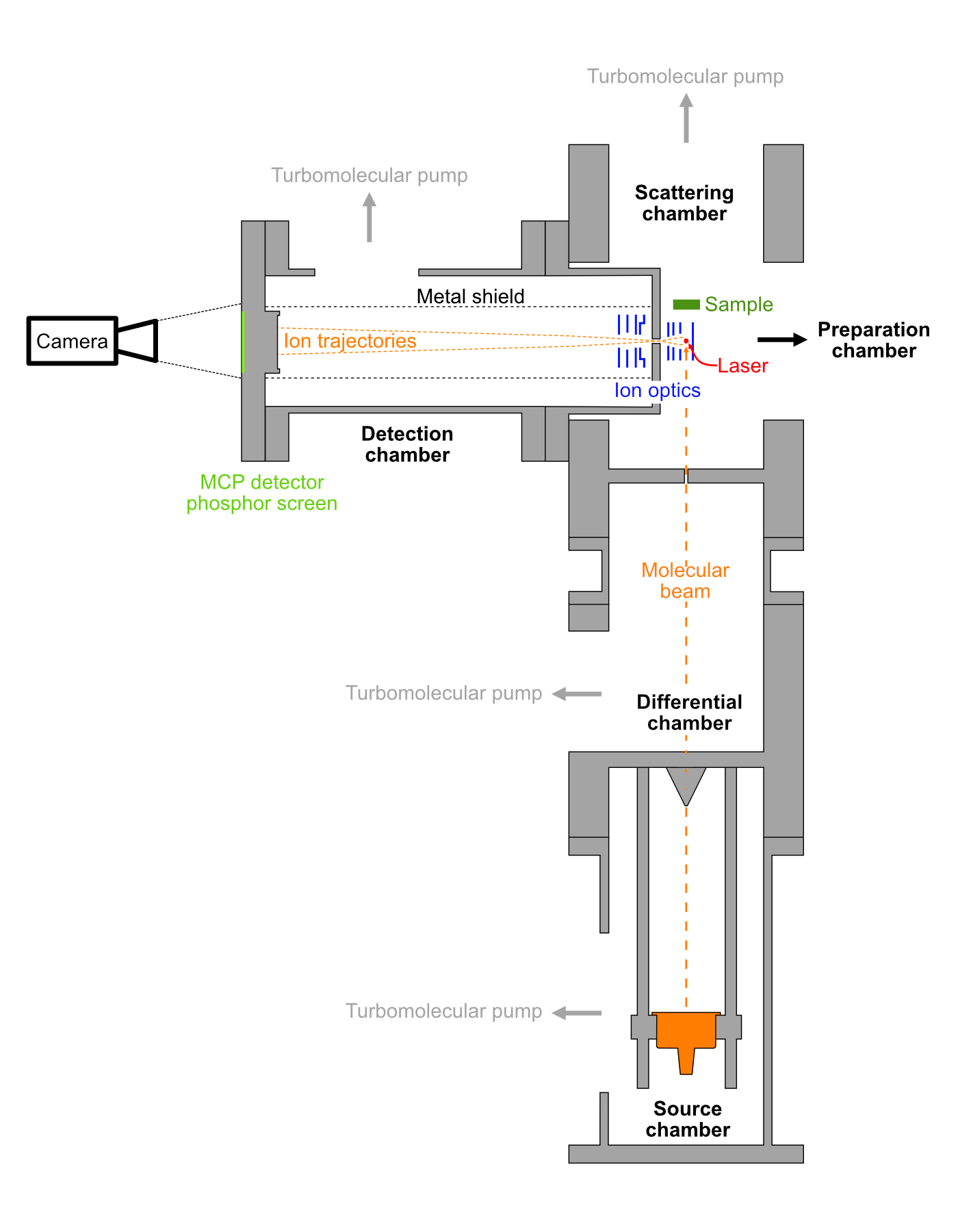}
    \caption{Schematic diagram of the NAP-VMI apparatus.}
    \label{fig:Instrument}
\end{figure}

Figure \ref{fig:Instrument} shows the schematic diagram of the apparatus. It consists of a molecular beam source chamber, scattering chamber, detection chamber, sample preparation chamber and laser system. The scattering  and detector chambers are pumped by  300\, l/s and 80\, l/s turbomolecular pumps, respectively. A top hat shaped flange is installed between them. A 3\, mm aperture on the top hat separates the chambers, providing differential pumping and allowing pressure differences of a factor of 1000 to be maintained.  Electrodes on either side of the aperture provide the fields needed for velocity mapping.

The instrument is controlled by home made Python software which controls the camera, BNC delay generator, and laser wavelength. Different types of experiment can be performed allowing integration over the speed distribution of the molecular beam or kinetic scans, where an image is saved for each MB-laser delay, needed to perform Velocity Resolved Kinetics measurements\cite{harding2017ion}. Image manipulation is performed using the PyAbel package\cite{PyAbel} in Python. 

\subsection*{B. Molecular beam}

The molecular beam source consists of two chambers which are separated by a skimmer (Beam Dynamics Inc., Model 1 with 2 mm opening). Source chamber and differential chamber are pumped by 350 l/s and 90 l/s turbomolecular pump, respectively.  A pulsed valve (Amsterdam Piezo Valve) is mounted in the source chamber 
which runs at 10 Hz repetition rate with a pulse width between 20-100 $\mu$s. The molecular beam source is mounted below the scattering chamber, separated by a 1 mm aperture. With the pulsed valve in operation the pressure of the source chamber increases to $1\times10^{-5}$ mbar, the pressure of the differential chamber increases to $1\times10^{-7}$ mbar, and the pressure of the scattering chamber is unchanged.

\subsection*{C. Detector chamber}

The detector chamber is pumped by a 70 l/s turbomolecular pump. A 40 mm Z-stack MCP and P43 phosphor screen (Photek) is used. 
The ion cloud is sliced by applying a short pulse of -500V to the front MCP. Light from the phosphor screen is then captured by a CCD camera (FLIR, 1.3 megapixel) with an 8\,mm lens.
The ion flight distance is 260\,mm. A metal grid is used to shield the field free flight region from the high voltage feedthroughs in the detector chamber. 

\subsection*{D. VMI ion optics}

In order to minimize scattering with background gas under NAP conditions the distance from the surface to the ionization region must be kept small. Consequently, the electrodes for the VMI optics also need to be small. The small size of the ion optics makes it more difficult to achieve high velocity resolution imaging compared to a larger set-up; for given voltages, the field gradients are larger, the ionization volume of the laser is a larger fraction of the extraction region, and mechanical imperfections are more important. The weak extraction fields we use should increase the volume over which VMI conditions are met.\cite{Reid2013Validation} 

\begin{figure}[h]
    \centering
    \includegraphics[scale=0.1]{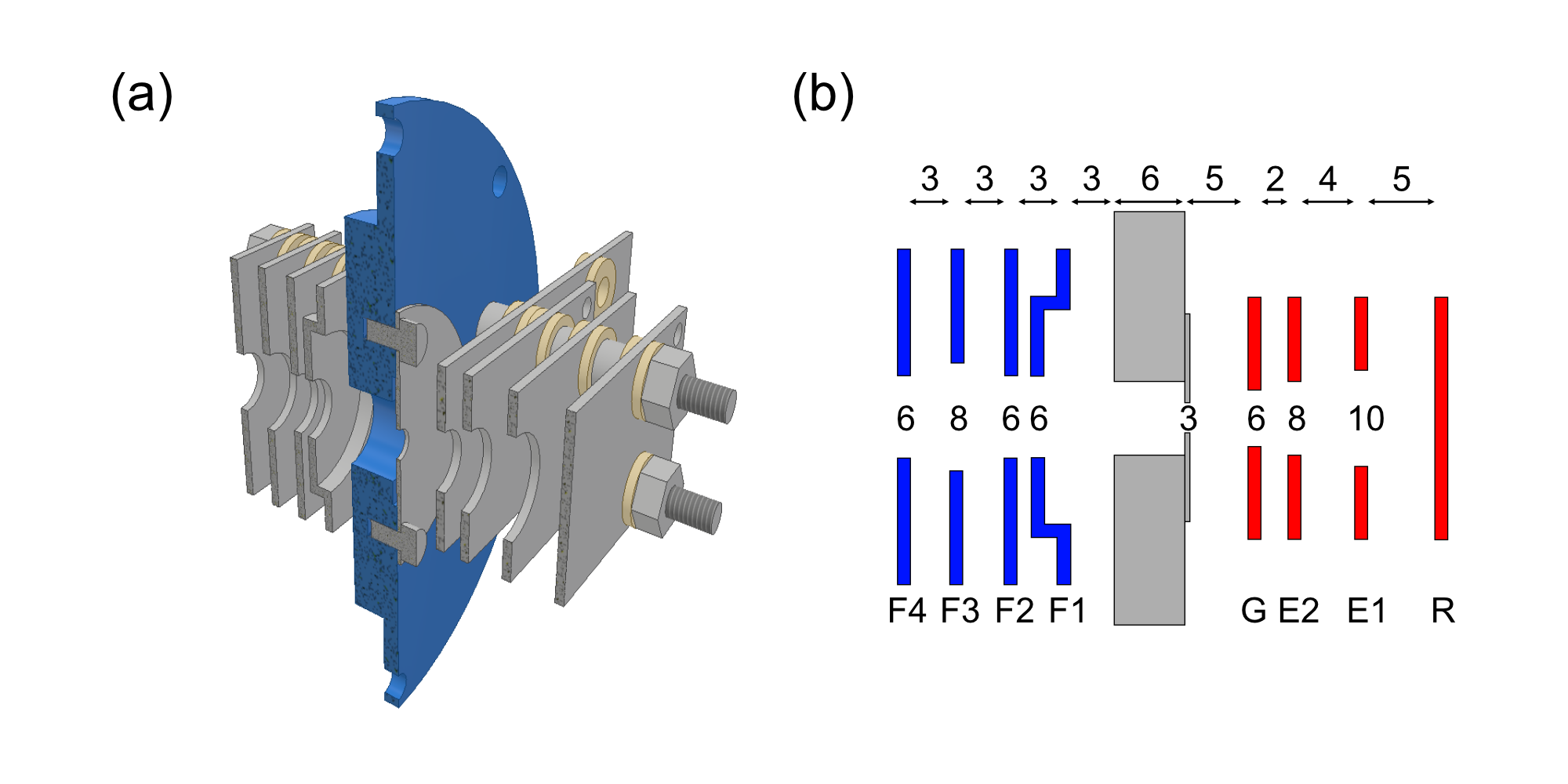}
    \caption{(a) The cross section and (b) the construction diagram of the ion optics.}
    \label{fig:Ionoptics}
\end{figure}

\begin{figure}[h]
    \centering
    \includegraphics[scale=0.1]{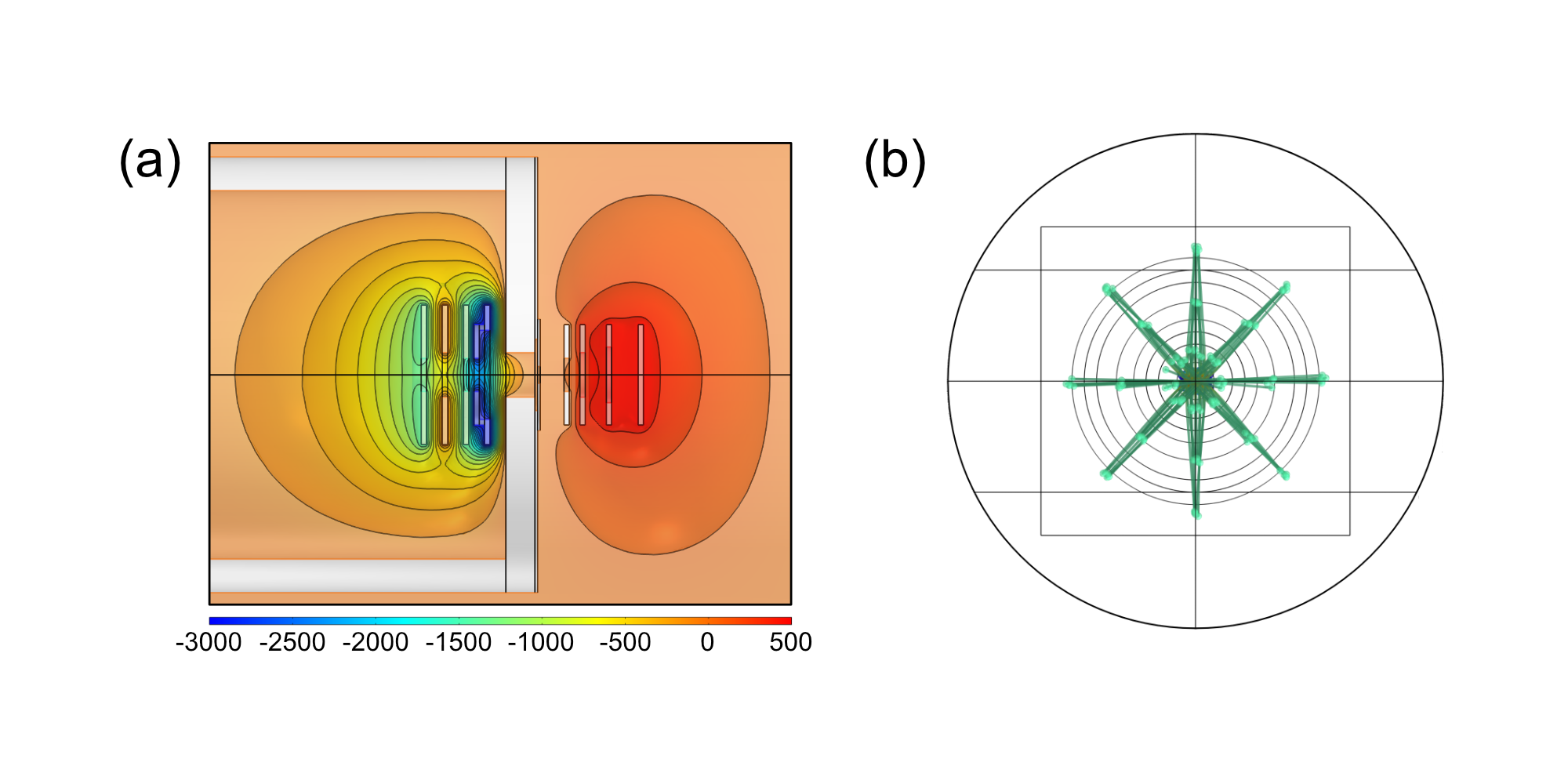}
    \caption{Simulation of (a) electric field and (b) ion trajectory for the VMI set-up}
    \label{fig:Simulation}
\end{figure}

Figure \ref{fig:Ionoptics} shows the VMI ion optics setup. It consists of eight 1 mm thick stainless steel electrodes. The first set of four electrodes (R, E1, E2, G) are designed to produce a hard focus of the ions at the aperture on the top hat. The second set of four electrodes (F1, F2, F3, F4) function as an acceleration region and an Einzel lens that fine tunes the focusing. Electrode R is a solid rectangle (85\,mm x 18\,mm), electrodes E1, E2, and E3 are rectangular plate (52 mm x 18 mm) with 10 mm, 8 mm and 6 mm hole in the center, respectively. The distance between electrodes R, E1, E2, G  are 5 mm, 4 mm and 2 mm; the distance between electrode G  and the top hat flange is 5 mm and the wall thickness of the top hat is 6 mm. F1 electrode is a square plate (25 mm $\times$ 25 mm) with concentric circles extrusion (18\,mm outer diameter and 16\,mm inner diameter) forming a shape of a top hat and have a 6\,mm hole in the center. Electrodes F2, F3 and F4  are square plate (25 mm x 25 mm) with 6\,mm, 8\,mm and 6\,mm hole in the center, respectively. The distance between the top hat flange, F1, F2, F3 and F4 are 3\,mm. Figure \ref{fig:Simulation} shows the electric field and ion trajectory simulation of \ce{N$_2^+$} ions with kinetic energy of 0.03 eV, 0.27 eV and 0.75 eV in COMSOL Multiphysics 5.4. Ions start from the center of R and E1 electrodes and 1 and 2 mm  along the laser propagation direction. Eight initial directions  parallel to the imaging plane with 45 degrees spacing were considered. The simulated voltages on R, E1, E2, G, F1, F2, F3 and F4 are 500\,V, 480\,V, 40\,V0, 0\,V, -3000\,V, -1500\,V, 0\,V, -1500\,V, respectively.
The ion optics also have a DC-slicing effect,\cite{townsend2003direct, lin2003application} spreading the arrival time at the detector for particles depending on their motion in the z-axis (towards the detector). Using fast gated MCPs allows only the center section of the image to be recorded, removing the need for an inverse-Abel transform.
For surface scattering experiments there is also a laser slicing effect\cite{TONOKURA1994Slicing} due to the distance from the scattering surface to the ionization point  (typically 14 mm). 
The out-of-plane angular acceptance  is mainly determined by the size of the molecular beam at the surface and we estimate that for the ca. 3\,mm beam only molecules within $\pm 15^{\circ}$ of the scattering plane parallel to the electrodes and detector will be ionized and these molecules will have apparent in-plane velocities up to 5\% too low. 
For most surface scattering experiments this is acceptable, but it could probably be improved using image analysis procedures similar to the FINA methods developed by Suits and co-workers\cite{FINA} 

A second mode of operation, using only the first electrode stack as 3-electrode VMI is also possible, despite the small size of the aperture. This mode is very similar to the optics described by Long et al. for PEI from a liquid microjet. 

The aperture has the additional benefit of effectively shielding the detector from light in the scattering chamber, which can come from scattering of the ionization laser, emission from the sample heater, or ionization filaments. 

\subsection*{E. Laser system}

The ionization laser is generated by doubled Nd:YAG laser (InnoLas) pumped dye laser (Radiant Dyes) The DCM dye fundamental around 609 nm is frequency tripled with two BBO crystals. The laser operates at a repetition rate of 10 Hz. A variable wave plate controls the laser beam polarization and it is focused in the center of the VMI optics by a 200 mm focal length lens.

\subsection*{F. Preparation chamber}

The preparation chamber is pumped by a 400 l/s turbomolecular pump with a base pressure below $2\times10^{-9}$ mbar. It contains a sputter gun (PREVAC), Auger Electron Spectrometer (RBD Instruments) and sample stage. The sample manipulator is controlled by UHV compatible linear translation and rotation stages (Arun Microelectronics Ltd.) allowing the sample to be located between sputtering, Auger and scattering regions. Prior to surface scattering experiments, the Pd(110) crystal was cleaned by argon ion sputtering for 5 min and annealed at 1000 K. Surface cleanliness was checked by Auger spectroscopy.

\section*{Results and Discussion}

\subsection*{A. \ce{N2O} photodissociation}

\begin{figure}[h]
    \centering
    \includegraphics[scale=0.1]{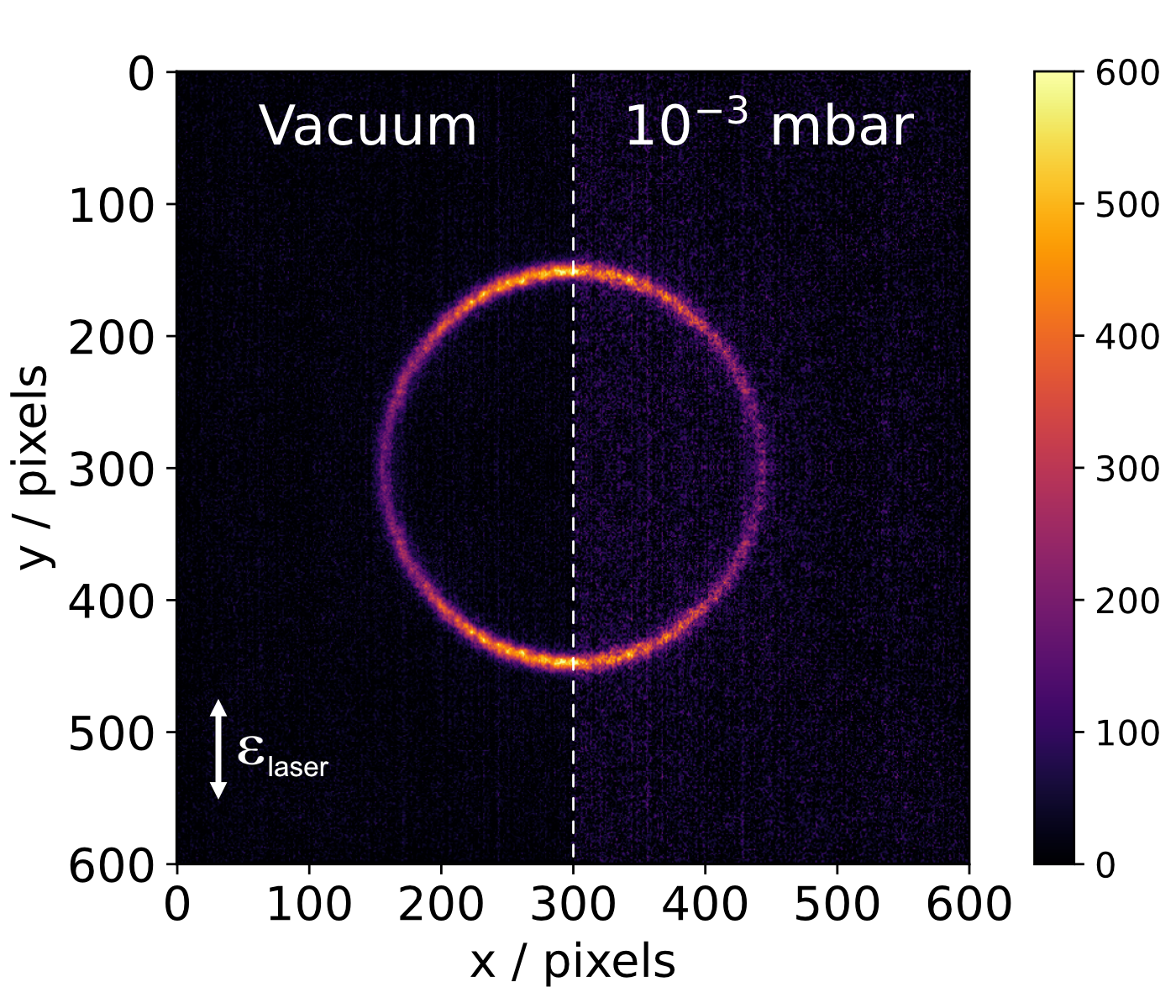}
    \caption{DC sliced images of N$_2$ (J=74) fragments from N$_2$O dissociation under high vacuum: $5\times10^{-9}$ (left panel) and near-ambient pressure: ($1\times10^{-3}$ mbar of Ar(right panel). The vertical arrow indicates the laser polarization ($\varepsilon_{laser}$) direction. 
    }
    \label{fig:N2OVMI}
\end{figure}

\begin{figure}[h]
    \centering
    \includegraphics[scale=0.125]{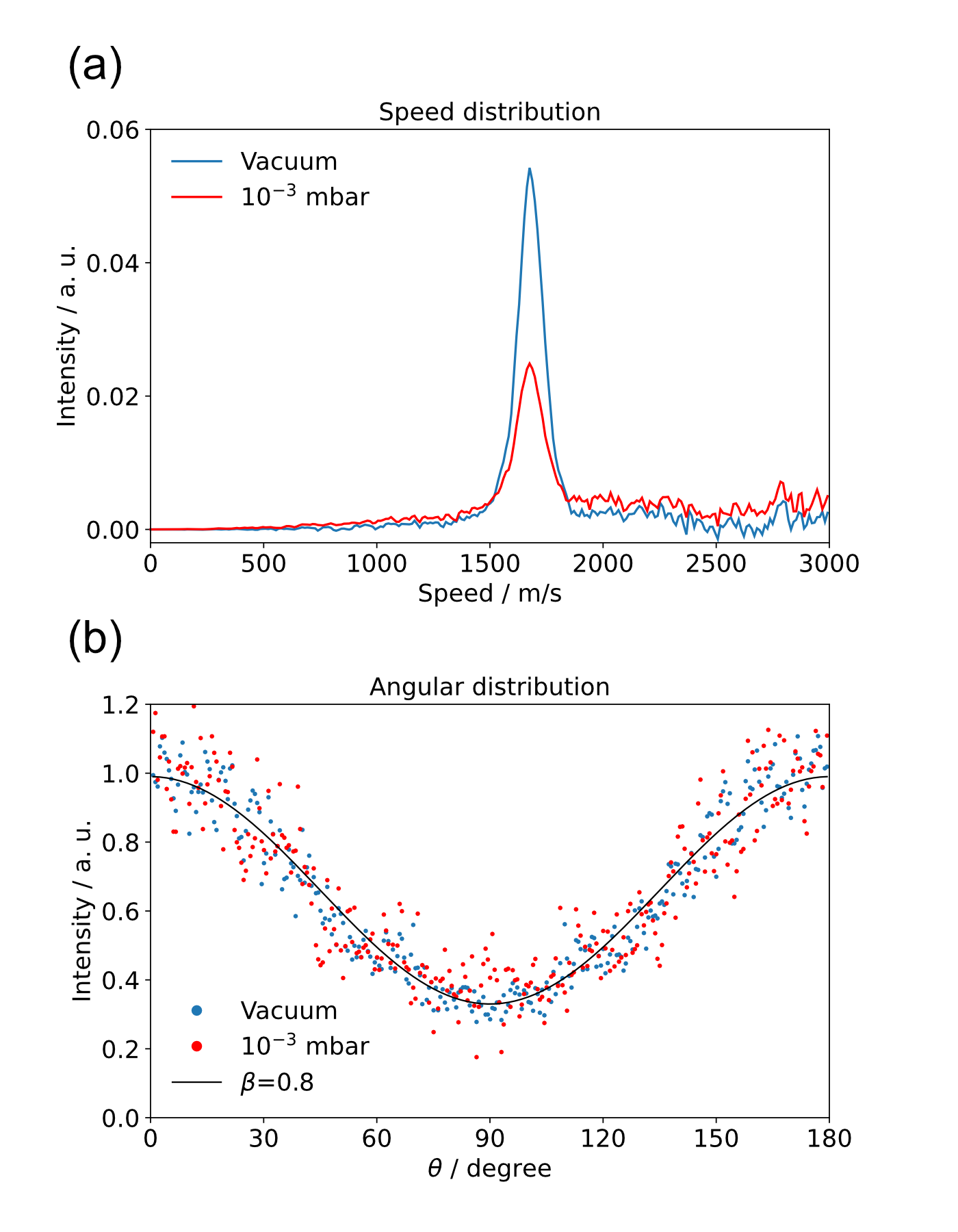}
    \caption{(a) Speed distribution and (b) angular distribution of \ce{N2} fragments obtained from Figure \ref{fig:N2OVMI}}
    \label{fig:N2Oanalysis}
\end{figure}

\begin{figure}[h]
    \centering
    \includegraphics[scale=0.125]{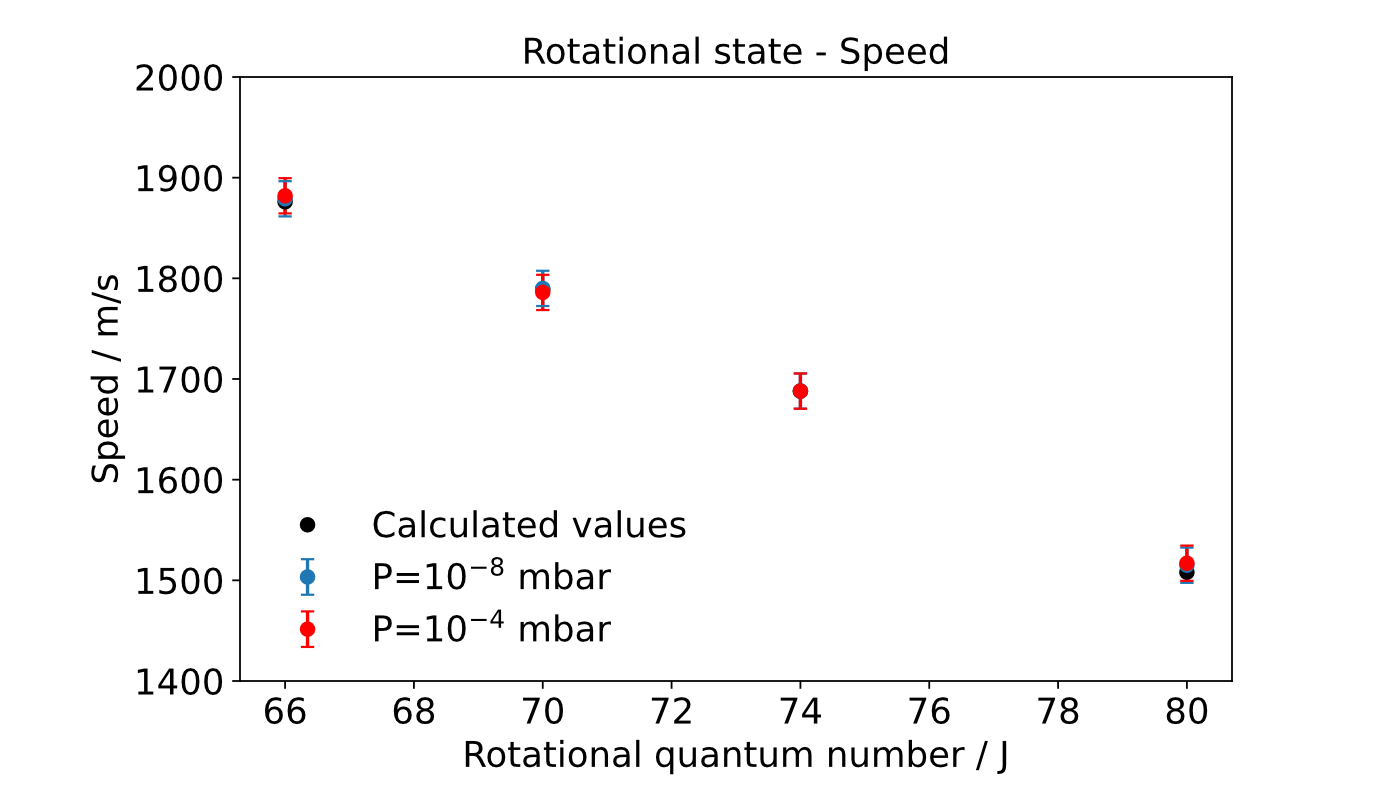}
    \caption{Plot of mean recoil speed for different final-J states under vacuum and medium pressure($ 1\times10^{-4} $ mbar of argon). The error bar was estimated to be 3 pixels($\sim$35 m/s).}
    \label{fig:Speed_J}
\end{figure}

To test and demonstrate the capabilities of the NAP-VMI optics, we have studied the well characterized photodissociation of \ce{N$_2$O}. Photodissociation of \ce{N$_2$O} around 203\,nm produces highly rotationally excited \ce{N$_2$} fragments which can be detected state selectively using (2+1) REMPI.\cite{hanisco1993state, suzuki1996evidence, neyer1999photodissociation, nishide2004photodissociation, kawamata2006photodissociation}  
The different final rotational states have significantly different translational kinetic energies, providing a good system to calibrate the new imaging optics. 
The optimized voltages of ion optics are experimentally set at 500\,V, 480\,V, 400\,V, 0\,V, -3000\,V, -1000\,V, -100\,V, -1000\,V 
for R, E1, E2, G, F1, F2, F3 and F4 electrodes, respectively. Figure \ref{fig:N2OVMI} shows a DC sliced velocity map image of N$_2$ (J=74) fragments from \ce{N$_2$O} dissociation under high vacuum ($5\times10^{-9}$ mbar: left panel) and near-ambient pressure ($1\times10^{-3}$ mbar of Ar: right panel). Figure \ref{fig:N2Oanalysis} shows the speed distribution obtained from Figure \ref{fig:N2OVMI}, the curve was used to calibrate the detector, yielding a value of 11.5 m/s per pixel. The full width at half maximum height of the peak is around 8\%. Compared to other imaging systems designed for photodissociation experiments\cite{nishide2004photodissociation} the speed resolution is not very high, and we are unable to resolve dissociation from different initial vibrational states of \ce{N2O}. This is presumably due to a combination of the small size of the optics and the fact that the molecular beam is parallel to the detector plane.  
The \ce{N2} speeds measured for different final rotational states are shown in Figure \ref{fig:Speed_J}. Comparison with calculated and published values\cite{nishide2004photodissociation, kawamata2006photodissociation} for other final-J states shows good agreement.  The measured angular distributions are also consistent with previous reports.\cite{neyer1999photodissociation}
At near-ambient pressure, collisions with the background Ar reduced the total signal and broadened the speed distribution (by factors around 3 in this case) but the useful, dynamical, information is still clearly visible.  

\subsection*{B. \ce{N2} scattering from the Pd(110) surface}

\begin{figure}[h]
    \centering
    \includegraphics[scale=0.125]{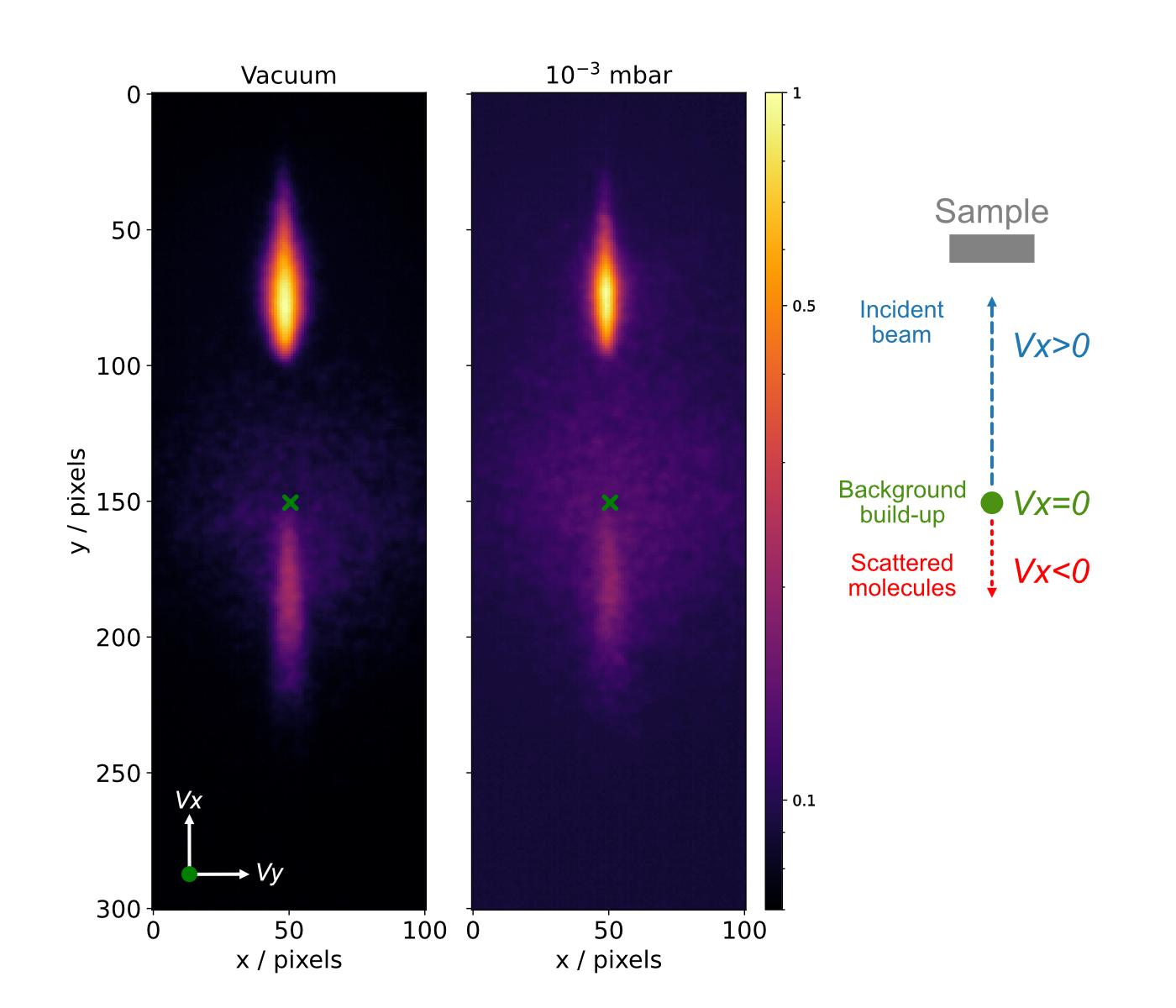}
    \caption{Velocity-map images for \ce{N2} scattering from Pd(110) surface (T$_s$ = 300 K) under vacuum and $1\times10^{-3}$ mbar of argon. The schematic illustration of sample, incident beam, background build-up, and scattered molecules are shown next to the images.}
    \label{fig:N2VMI}
\end{figure}

\begin{figure}[h]
    \centering
    \includegraphics[scale=0.125]{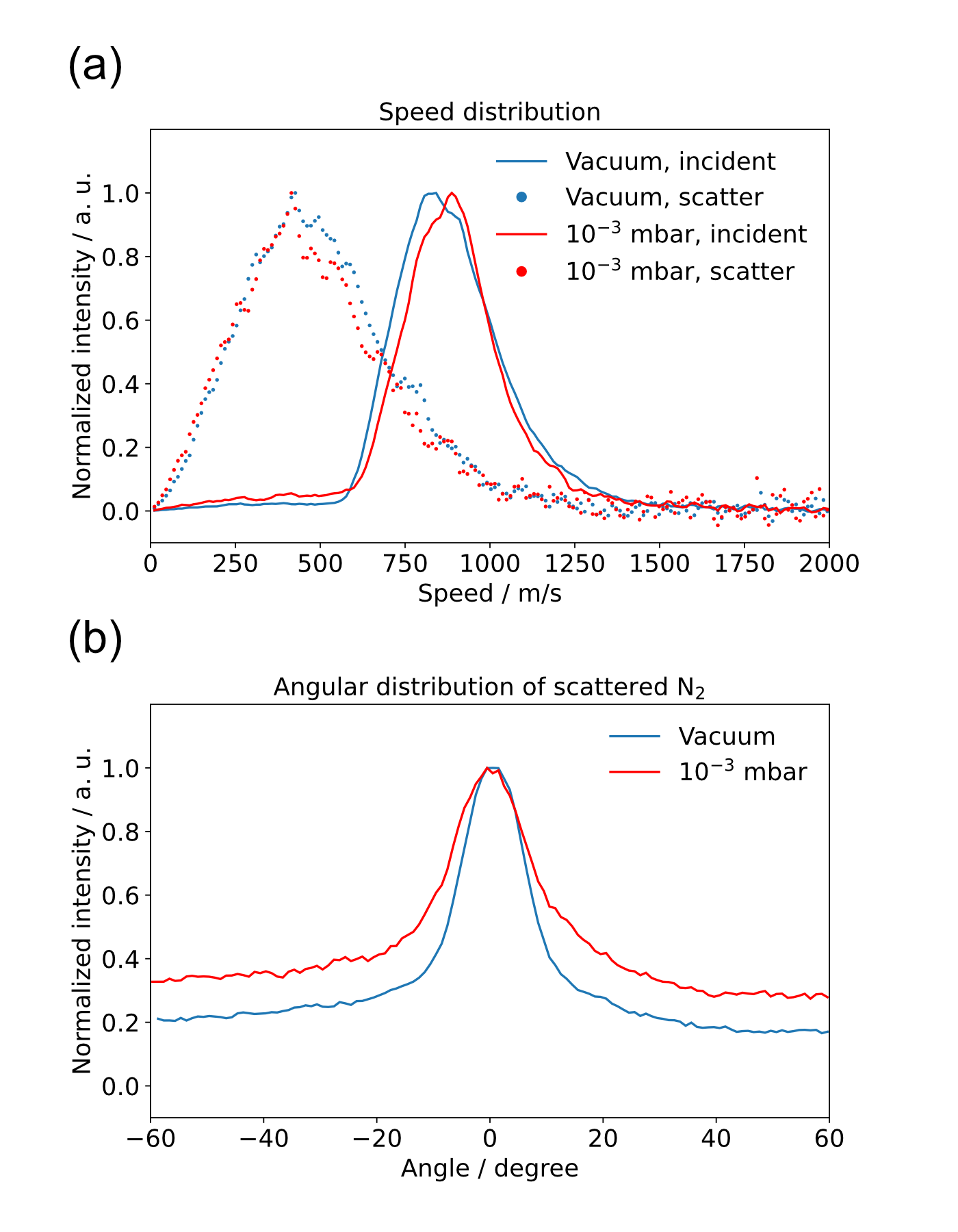}
    \caption{(a) Speed distributions from VMI images for incident beam and scattered molecules, and (b) angular distribution for scattered \ce{N2}.}
    \label{fig:N2analysis}
\end{figure}

We have tested the ability of the NAP-VMI optics to probe surface scattering by measuring  inelastic scattering of \ce{N2} from Pd(110) at 300\,K under  ultrahigh vacuum and in a near-ambient environment in argon. Tthe sample was placed 5\,mm from the edge of the ion optics, making a 14 mm distance from sample to ionization center. 
We choose this distance as it reduces perturbations to the imaging fields due to the presence of the surface. 
The ion optics are experimentally set at 500, 480, 400, 0, -3000, -1000, -75, -1000 V for R, E1, E2, G, F1, F2, F3 and F4 electrodes, respectively. \ce{N2} molecules are detected state selectively using (2+1) REMPI.\cite{hanisco1991resonantly} Figure \ref{fig:N2VMI} shows the velocity map images of \ce{N2} scattering from Pd(110) under vacuum and near-ambient environment ($1\times10^{-3}$ mbar of argon). The green cross in the image center represents the thermal background of \ce{N2} molecules, which we also use as a zero-velocity point; the upper signal ($V_x>0$) is the incident beam which is travelling upward (approaching surface), and the lower signal ($V_x<0$) belongs to scattering molecules which are travelling downward (leaving surface) after scattering. The images were recorded while scanning the delay between the pulse valve and the laser from 300 $\mu$s to 800 $\mu$s  with an interval of 10 us, so that the full velocity distribution of molecules in the beam is included. Figure \ref{fig:N2analysis} shows the speed distribution from VMI images, by integrating the radius over 20$^{\circ}$ sectors centered on the incident beam and the scattered molecules direction, and angular distribution of scattered molecules. Comparing the speed distribution between high vacuum and near-ambient environment, no significant difference is found. The root mean square speed for the incident beam is 910 m/s (0.12 eV) and scattered \ce{N2} is 535 m/s (0.042 eV), the scattered \ce{N2} has lost around 65\% of its kinetic energy. The observed energy loss is in good agreement with the Baule limit,\cite{baule1914theoretische} where the final translational energy is predicted by a hard cube model. The final translational energy $\langle E_f \rangle $, is described as $\langle E_f \rangle = \langle E_i \rangle \left( \dfrac{m_{Pd}-m_{N_2}}{m_{Pd}+m_{N_2}} \right)^2  $, and it predicts a 66\% energy loss for \ce{N2} scattering from Pd surface. The angular distribution for scattered \ce{N2} is broader in the near-ambient argon environment, which we attribute to collisions with Ar atoms.

\section*{Conclusion}
We have demonstrated a new instrument capable of studying  molecular beam surface scattering under near-ambient pressure, providing a new tool to help close the `pressure gap' in the studies of catalytic reaction. The design is based on a velocity-map image set--up that uses two sets of electrodes to focus the ions at an aperture. By putting the aperture between scattering and detector chamber, we created a differential pumping, and allow us operate under near-ambient pressure in the scattering chamber without damaging the detector. Higher pressures could be achieved by using a smaller aperture and adding additional differential pumping stages. The speed resolution is not very high, due to the compact design of the electrodes, but it reduces the number of background collision and preserves the dynamical scattering information about the reaction. This technique may be applied in a range of other areas where higher pressures are either interesting or unavoidable, e.g. liquid jets and gas-liquid interfaces 

\section*{Acknowledgment}
We acknowledge the financial support of the Swedish Foundation for Strategic Research (SSF) (ITM17-0236). We thank Henrik \"{O}str\"{o}m for the loan of the Pd(110) crystal. DJH thanks Alice Schmidt-May for carefully reading the manuscript.

\bibliography{references}

\end{document}